\documentclass[runningheads]{llncs}
\usepackage[T1]{fontenc}
\usepackage{graphicx}
\usepackage{hyperref}
\usepackage{tabularx, array}

\begin{document}

\title{GUARD-CAN: Graph-Understanding and Recurrent Architecture for CAN Anomaly Detection}
\titlerunning{{GUARD-CAN}}

\author{Hyeong Seon Kim \and Huy Kang Kim}
\authorrunning{H. S. Kim and H. K. Kim}
\institute{School of Cybersecurity, Korea University, Republic of Korea\\
\email{\{ccloseon, cenda\}@korea.ac.kr}}
\maketitle

\begin{abstract}
Modern in-vehicle networks face various cyber threats due to the lack of encryption and authentication in the Controller Area Network (CAN). To address this security issue, this paper presents \textbf{GUARD-CAN}, an anomaly detection framework that combines graph-based representation learning with time-series modeling. \textbf{GUARD-CAN} splits CAN messages into fixed-length windows and converts each window into a graph that preserves message order. To detect anomalies in the time-aware and structure-aware context at the same window, \textbf{GUARD-CAN} takes advantage of the overcomplete Autoencoder (AE) and Graph Convolutional Network (GCN) to generate graph embedding vectors. The model groups these vectors into sequences and feeds them into the Gated Recurrent Unit (GRU) to detect temporal anomaly patterns across the graphs. \textbf{GUARD-CAN} performs anomaly detection at both the sequence level and the window level, and this allows multi-perspective performance evaluation. The model also verifies the importance of window size selection through an analysis based on Shannon entropy. As a result, \textbf{GUARD-CAN} shows that the proposed model detects four types of CAN attacks (flooding, fuzzing, replay and spoofing attacks) effectively without relying on complex feature engineering.

\end{abstract}

\section{Introduction}
Recent vehicle systems focus on automation, electrification, and connectivity. This trend increases the complexity and importance of in-vehicle networks. A representative example is the Controller Area Network (CAN)~\cite{ref_CAN}, which is widely used based on its real-time performance and lightweight protocol.

However, from a security perspective, CAN lacks encryption and authentication functions, making it vulnerable to various cyberattacks~\cite{ref_CAN_vul}. In July 2015, Miller~\textit{et al.} disclosed a remote hacking case involving a Jeep Cherokee~\cite{ref_miller}. They showed that a real vehicle could be controlled through network intrusion, which raised strong awareness of vehicle security in both academia and industry. In addition, Tencent Keen Security Lab proved remote hacking by injecting malicious messages into the CAN bus of Tesla Model S and Model X vehicles through wireless (Wi-Fi/cellular) networks~\cite{ref_freefall}. As vehicles are transforming from simple means of transportation into complex cyber-physical systems, the importance of CAN-based anomaly detection and real-time security systems continues to grow. Given these trends, various intrusion detection systems (IDSs) have been proposed to detect security threats in CAN. Statistical, machine learning-based, and deep learning-based methods have been used, but existing studies often require domain knowledge of the CAN bus and have the limitation of requiring additional feature engineering.

This paper proposes \textbf{GUARD-CAN}, which detects temporal anomalous patterns in CAN bus attacks. \textbf{GUARD-CAN} divides CAN messages into fixed-size windows and converts each window into a graph. It combines overcomplete Autoencoder (AE) and Graph Convolutional Network (GCN) to encode patterns embedded in the data. The encoded graph embedding vectors are formed into a sequence and input into Gated Recurrent Unit (GRU) to learn and detect anomalous patterns in the temporal context. The main contributions of this study are as follows:

\begin{itemize}
    \item \textbf{Arbitration ID independent algorithm}: \textbf{GUARD-CAN} only uses DLC and Data field as graph node features. That means \textbf{GUARD-CAN} does not need a full description of DBC, which is regarded as a private intellectual property.
    \item \textbf{Structure-aware graph embedding with GCN}: We combine overcomplete AE with GCN to encode each graph into a graph embedding vector. This process extracts structural and temporal patterns from the data.
    \item \textbf{Temporal context learning with GRU over graph sequences}: We group graph embedding vectors by sequence length and use them as input to GRU. This enables anomaly detection that reflects temporal characteristics across graphs.
\end{itemize}

\section{Background}
\subsection{Controller Area Network (CAN)}
CAN is an in-vehicle communication protocol widely used in modern vehicles. Based on its durability and efficiency, it supports real-time data transmission and reception between various Electronic Control Units (ECUs)~\cite{ref_CAN}. However, CAN lacks message authentication and does not apply message encryption. Furthermore, as all ECUs are connected to a common network and communicate via broadcast, they are exposed to many security threats~\cite{ref_CAN_vul}. The dataset used in this study~\cite{ref_dataset2} includes a total of four types of CAN bus attacks, which are as follows:
\begin{itemize}
    \item \textbf{Flooding}: Transmits an excessive number of messages to overwhelm the CAN bus and exhaust its communication bandwidth.
    \item \textbf{Fuzzing}: Randomly injects arbitrary messages to cause system malfunctions.
    \item \textbf{Replay}: Reuses previously recorded normal CAN messages to disguise malicious activity as legitimate traffic.
    \item \textbf{Spoofing}: Alters specific CAN IDs and data based on traffic analysis to manipulate vehicle functions.
\end{itemize}

\subsection{Graph Convolutional Network (GCN)}
GCN~\cite{GCN} is a neural network model designed to effectively process graph data with irregular structures. It extends the concept of Convolutional Neural Networks, which are primarily used in image processing, to the graph domain. GCN integrates not only the features of each node but also the features of neighboring nodes and their connection structure. GCN performs convolution operations on the graph at each layer and updates the node feature representations accordingly. In particular, GCN learns meaningful patterns from data without complex feature engineering by aggregating information from directly connected neighboring nodes.

In this study, we combine overcomplete AE and GCN. This model extracts graph embedding vectors that effectively capture the structural and temporal features in CAN bus graph data. Thereby, GCN contributes to the stability and performance improvement of anomaly detection.

\section{Related Work}
Existing IDS for CAN studies can be classified into statistical anomaly detection, machine learning-based detection, and deep learning-based detection. Song~\textit{et al.}~\cite{song-timeinterval} proposed a lightweight IDS for CAN by analyzing the time interval between CAN messages. Lee~\textit{et al.}~\cite{OTIDS} proposed an IDS that detects attacks on CAN networks through response pattern analysis based on remote frames. The authors used the offset and time interval between request and response messages as key features. Song~\textit{et al.}~\cite{song-DCNN} proposed an IDS for CAN based on a deep convolutional neural network with a lightweight Inception-ResNet structure. This method achieved high performance by extracting meaningful patterns from CAN traffic data without complex feature engineering.

Ye~\textit{et al.}~\cite{GDTIDS} proposed three graph-based features (time difference, betweenness centrality, and graph density) that can be applied to intrusion detection. They used these features to perform anomaly detection using Classification and Regression Trees. Song~\textit{et al.}~\cite{DGIDS} introduced DGIDS, which builds a dynamic graph based on the arrival order of messages. Instead of using a fixed-length window, they used the message cycle of a base ID as the window. The system runs both offline and online phases simultaneously and detects anomalies by using multiple features. Devnath proposed GCNIDS~\cite{GCNIDS}, a GCN-based IDS designed to detect mixed attacks in CAN bus data. The method constructs graphs using Arbitration IDs within each window and uses only the maximum in-degree and out-degree as node features. GCN extracts node embeddings and performs graph-level binary classification.

However, existing studies~\cite{song-timeinterval,OTIDS,GDTIDS,DGIDS} require domain knowledge of CAN bus data and may have limitations in learning complex patterns embedded in the data. This study considers these points and allows GCN to learn meaningful patterns without complex feature engineering. In addition, the existing method~\cite{GCNIDS} only analyzes graphs from individual time steps without considering temporal continuity. This study addresses this by using sequences to reflect temporal characteristics between graphs.

\section{Methodology}
\begin{figure}
\includegraphics[width=\textwidth]{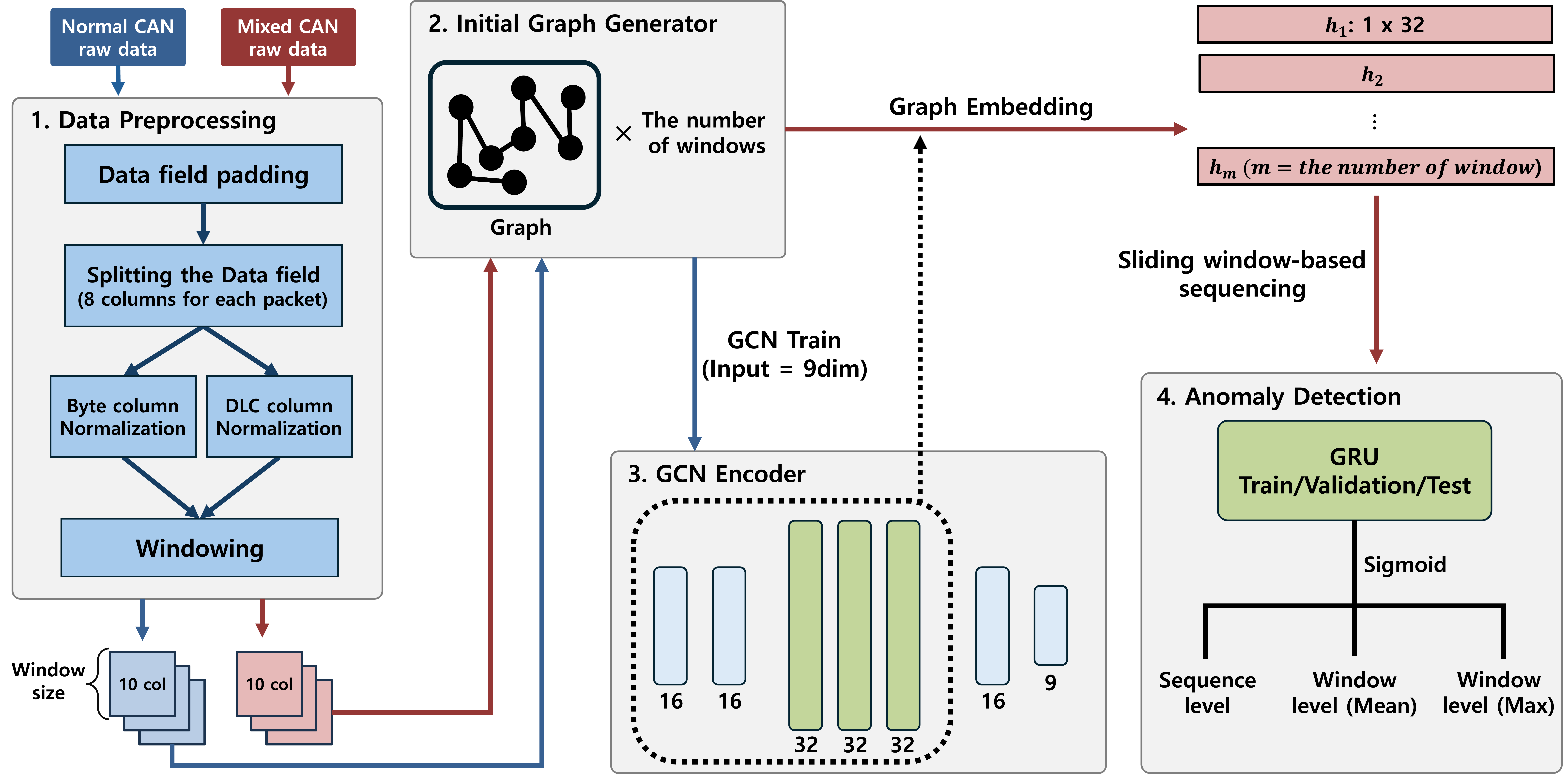}
\caption{Overview of \textbf{GUARD-CAN}.} \label{fig1}
\end{figure}

This section presents the overall methodology of \textbf{GUARD-CAN}. The proposed framework consists of four main steps, and the overall pipeline is illustrated in Fig.~\ref{fig1}.

\subsection{Data Preprocessing}
In this study, we applied a series of preprocessing steps to convert raw CAN data into a form suitable for model training. A raw CAN packet typically consists of the Timestamp, the Arbitration ID, the Data Length Code (DLC), and the Data field.

In the first step of preprocessing, we add 0x00 bytes to the end of the data field when its length is shorter than 8 bytes. After padding, we split each data field into eight columns from byte1 to byte8. Then, we convert each byte value from hexadecimal to decimal and normalize it to a real value in the range [0, 1]. The DLC value is normalized in the same manner. This normalization process mitigates the scale differences between the integer-based DLC field and the byte-based Data field. This aims to prevent potential instability in model training.

Following feature transformation, we segment the CAN messages into fixed-size windows using a non-overlapping windowing strategy based on a predefined window size. If the last window contains fewer packets than the specified window size, it is discarded. For example, with a window size of 100, a final window containing only 79 packets would not be used in training or evaluation. Simultaneously, we assign labels to each window. If any packet within a window is labeled as an attack, the entire window is marked as abnormal.

\subsection{Initial Graph Generator}
The Initial Graph Generator takes the preprocessed windows and converts each window into a graph. The model uses these graphs as input to GCN and learns the structural characteristics of the CAN data flow. It constructs edges by sequentially connecting packets based on their timestamps and preserves the temporal order of message transmissions within each window. This edge design helps the model capture the temporal dependencies among consecutive CAN messages.

The node features consist of the normalized DLC (\textit{DLC\_norm}) and the normalized values of the CAN payload (byte1–byte8) obtained during preprocessing. At this point, we convert the processed CAN payload values into integers before using them as the node features. This design prevents unnecessary increases in the dimensionality of GCN. With these features, the model can effectively identify anomalous data values that deviate from the normal message flow, thereby improving anomaly detection performance. By leveraging these two types of node-level features, \textbf{GUARD-CAN} eliminates the need for manual domain-specific feature engineering.

\subsection{GCN Encoder}
GCN takes each window-level graph as input and performs encoding to generate a graph embedding vector. \textbf{GUARD-CAN} uses only normal data to train GCN and applies the trained GCN to encode graphs that contain both normal and attack data. In this study, we combine overcomplete AE with GCN to improve its training effectiveness. \textbf{GUARD-CAN} does not rely on complex feature engineering. Instead, it extracts features that reflect both temporal and structural characteristics within the graph using reconstruction loss from the AE. By using this structure, the model produces more generalized representations and improves training stability compared to using GCN alone.

First, we use the node features (\textit{DLC\_norm}, binarized byte1 to byte8) as input to the AE encoder. The encoder consists of two linear layers and applies ReLU activation function after each layer. It takes a 9-dimensional input and embeds it into a 16-dimensional latent space. AE encoder transforms the input features into expressive vectors. During this process, it extracts a latent vector that reflects the characteristics of each node and helps GCN learn more effectively.

The embedding vectors extracted from the AE encoder are used as inputs to GCN. GCN consists of three GCNConv layers, each followed by ReLU. All three layers output 32-dimensional embedding vectors. Through the graph structure, each node updates its latent vector by aggregating features from itself and its neighboring nodes. Additionally, the edge connections based on temporal order allow GCN to indirectly capture sequential patterns. This process enables the generation of graph-level representations that incorporate the overall context of the graph.

Finally, AE decoder takes the graph embedding vector from GCN as input and reconstructs it into the original input feature space. It consists of two linear layers. The first layer converts a 32-dimensional to 16-dimensional. After applying ReLU, the final layer transforms the 16-dimensional vector into the original input dimension of 9. This process trains the model to reconstruct the original node features. The model evaluates reconstruction accuracy using Mean Squared Error (MSE) loss. 

When embedding mixed data, the model loads the saved parameters and uses only AE encoder and GCN. AE encoder maps node features into 16-dimensional vectors, and GCN updates these vectors based on graph context. Then it applies global mean pooling to generate a 1×32 graph embedding vector for each window. The anomaly detection model takes these graph embedding vectors as input.

\subsection{Anomaly Detection}
\subsubsection{Creating Sequences}
This study aims to reflect the temporal characteristics and patterns among graph embedding vectors in anomaly detection. For this purpose, we organize graph embedding vectors into fixed-length sequences and feed them into the GRU-based anomaly detection model. \textbf{GUARD-CAN} constructs the sequences using a method similar to a sliding window. For example, when the sequence length is 3, the sequences are generated as follows: $S_0 = [h_0, h_1, h_2]$, $S_1 = [h_1, h_2, h_3]$, $\cdots$, $S_n = [h_{(m-2)}, h_{(m-1)}, h_m]$, where $S_n$ represents the ${n}$-th sequence and $h_m$ denotes the graph embedding vector encoded by GCN. The variable $m$ indicates the total number of windows. The model assigns a label to each sequence based on the labels of all the graphs that form the sequence. If the sequence includes even one graph labeled as an attack, the model treats that sequence as anomalous.

\subsubsection{GRU Model Configuration}
The GRU model receives input in the form of [\textit{batch\_size}, \textit{sequence\_length}, \textit{feature\_dim}]. \textit{batch\_size} denotes the number of sequences; \textit{sequence\_length} indicates the number of windows in each sequence; and \textit{feature\_dim} represents the dimension of the graph embedding vector generated for each window. We set the \textit{feature\_dim} to 32 because the graph embedding vector extracted by GCN has a shape of 1×32.

The model uses two GRU layers to learn temporal dependencies within each sequence. We set the hidden dimension of GRU to 64 and apply a dropout rate of 0.3 to prevent overfitting. The GRU output passes through two fully connected (FC) layers to produce the final prediction. The first FC layer compresses the GRU output into a 32-dimensional representation, and the second FC layer generates binary classification probabilities for each sequence. We apply sigmoid activation function to obtain the final probability values.

This study evaluates anomaly detection performance at both the sequence-level and the window-level. Sequence-level detection captures anomalies by considering the temporal flow between graphs, while window-level detection focuses on identifying anomalies within each graph. In this study, optimization is performed using Binary Cross Entropy (BCE) loss based on the sequence-level predictions.

\subsubsection{Sequence-level Anomaly Detection}
A sequence consists of a set of graph embedding vectors, and we regard each sequence as an independent unit. The sequence-level anomaly probability output by GRU ranges between 0 and 1. We apply a threshold of 0.5 to determine whether an anomaly is detected. As a result, the final output of the sequence-level anomaly detection is returned as a binary value of 0 or 1.

\subsubsection{Window-level Anomaly Detection}
A window refers to a group of CAN packets, and each window forms a single graph. This study also evaluates anomaly detection performance at the window-level, which is a finer unit than the sequence. Through this approach, we analyze how the temporal characteristics between graphs affect detection performance. Window-level anomaly detection is performed based on the window-level anomaly probability output from the GRU model. Each sequence uses overlapping graph embedding vectors, so a single window can appear in multiple sequences. Therefore, we perform window-level anomaly detection using two approaches.

The first approach determines anomalies based on the average of all predicted values for each graph embedding vector $h$. The second approach determines anomalies based on the maximum predicted value for each $h$. For both approaches, the anomaly detection threshold is set to 0.5, and the final output for each window is a binary value of 0 or 1.

\section{Experiment}
This section evaluates the performance of \textbf{GUARD-CAN}. We first describe the dataset used in the experiments and then evaluate the performance based on changes in window size and sequence length. Experiments are performed on MS Windows 10, Intel(R) Core(TM) i7-10700K CPU @ 3.80 GHz, 64.0 GB RAM, NVIDIA GeForce RTX 3080 Ti (12GB). And we implement \textbf{GUARD-CAN} using Python 3.12.2, torch 2.4.1, and torch-geometric 2.6.1.

\subsection{Dataset}
In this study, we conducted experiments using the Car Hacking: Attack \& Defense Challenge 2020 dataset~\cite{ref_dataset2,ref_dataset1}, provided by HCRL, Korea University. GCN learns the temporal characteristics and graph structure of normal data during training. For anomaly detection, we use a dataset that contains both normal data and four types of attack data. This dataset was split into training, validation, and test sets in a 6:2:2 ratio. The composition of the raw dataset, including the normal and four attack types, is shown in Table~\ref{dataset-stats}.

\begin{table}
\caption{Distribution of normal and attack records by dataset.}
\label{dataset-stats}
\centering
\begin{tabular}{|p{3cm}|p{2cm}|p{5cm}|}
\hline
\centering Dataset & \centering Type & \centering Number of Records (Ratio) \tabularnewline
\hline
\centering GCN training   & \centering Normal    & \centering 179,346 (100\%) \tabularnewline
\centering Anomaly detection     & \centering Normal    & \centering 1,799,046 (89.92\%) \tabularnewline
\centering Anomaly detection                & \centering Flooding  & \centering 96,559 (4.83\%) \tabularnewline
\centering Anomaly detection                & \centering Fuzzing   & \centering 44,770 (2.24\%) \tabularnewline
\centering Anomaly detection                & \centering Replay    & \centering 37,869 (1.89\%) \tabularnewline
\centering Anomaly detection                & \centering Spoofing  & \centering 22,489 (1.12\%) \tabularnewline
\hline
\end{tabular}
\end{table}

\subsection{Window Size Analysis}
In this study, we divide each packet according to the designated window size during the data preprocessing step. Then, using the initial graph generator, we convert each window into a graph based on the order of messages. Therefore, the window size serves as a key parameter that defines the unit of the graph. If the window size is too small, the graph lacks sufficient information, which limits what GCN can learn. In contrast, if the window size is too large, the graph may include excessive and diluted information, which can negatively affect model training.

Therefore, we analyze the entropy of each window to determine an appropriate window size. In general, when the IDs appear more evenly within a window, the entropy value increases. This implies that the information is evenly distributed across the window. Selecting a proper window size helps GCN extract meaningful features and ultimately improves anomaly detection performance. To support this process, we perform an analysis using Shannon entropy about window sizes ranging from 10 to 400.

\begin{figure}
\centering
\includegraphics[width=0.7\textwidth]{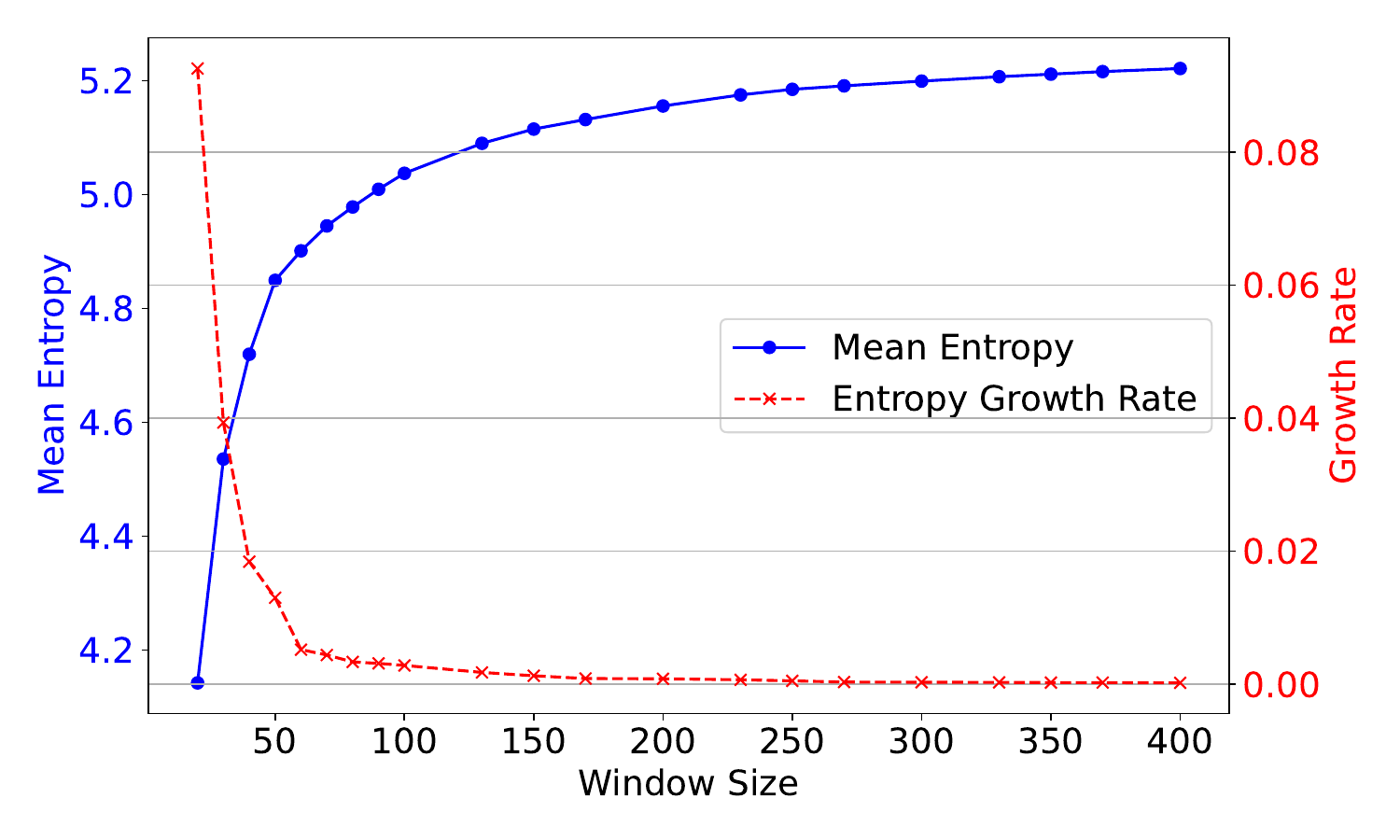}
\caption{Entropy average and growth rate by window size. This graph shows the values between window sizes 20 and 400, excluding window size 10 for which the growth rate cannot be calculated.} \label{fig2}
\end{figure}

We calculated the following statistical values for each window divided by the designated size: mean, median, minimum, maximum, and standard deviation of entropy. Based on these statistics, we also analyzed the entropy growth rate by window size. Fig.~\ref{fig2} shows the average entropy values and growth rate trends for each window size.

As shown in Fig.~\ref{fig2}, when the window size is less than 50, the entropy increases rapidly. In cases where the amount of information increases sharply, the graph may still lack enough information for GCN to learn effectively. In addition, it can be observed that for window sizes over 150, the growth rate nearly converges to zero. Therefore, this study conducted performance evaluations of the proposed model using five window sizes: 50, 75, 100, 125, and 150. In addition, the sequence length determines the temporal context that GRU can use to detect anomalies. Short sequences help capture local anomalies, and long sequences provide broader context but may dilute the influence of recent abnormal patterns. To evaluate performance under different levels of temporal dependency, we selected sequence lengths of 30, 50, 100, 120, and 150 for the experiments.  

\begin{figure}
\centering
\small
\includegraphics[width=\textwidth]{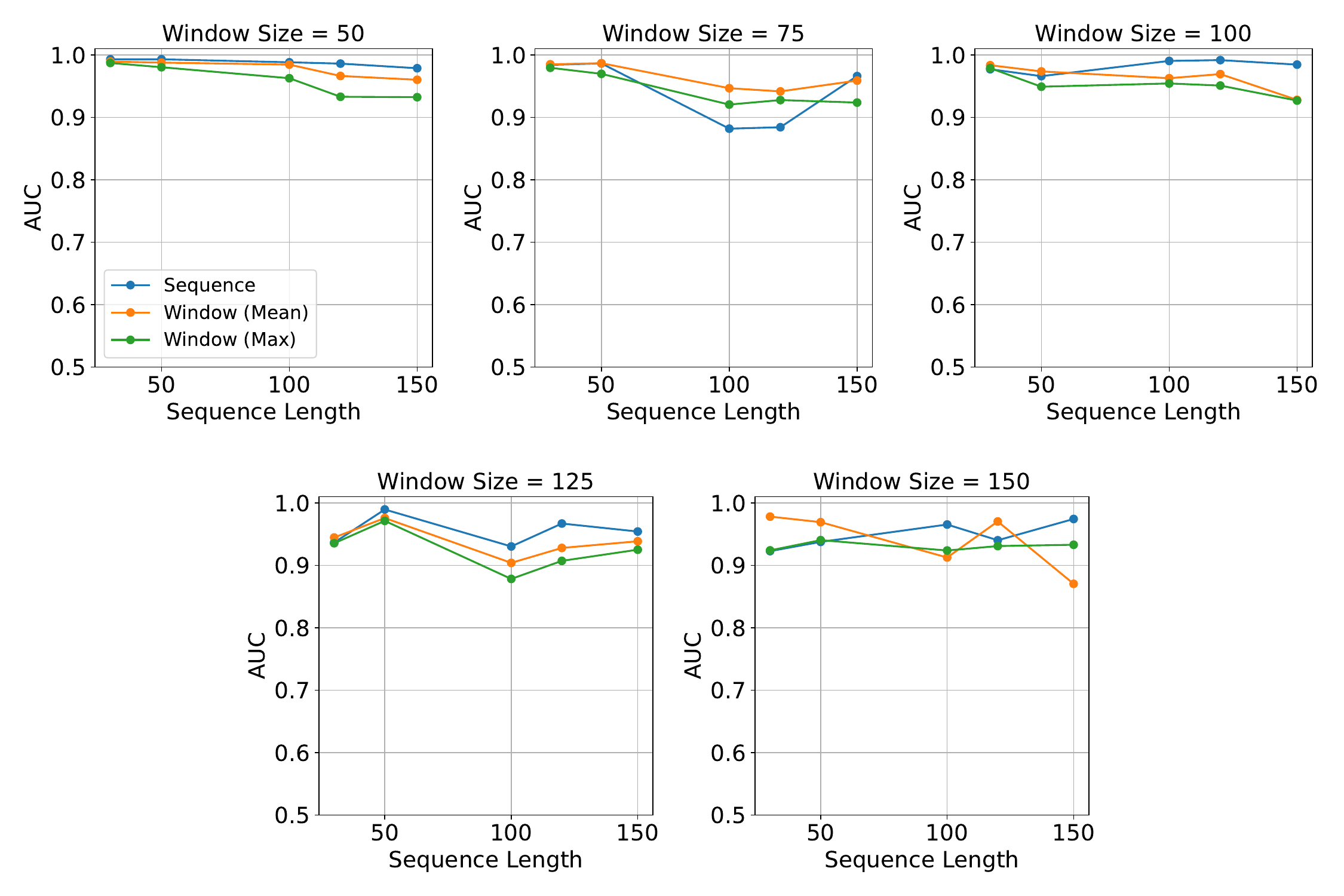}
\caption{AUC variation by sequence length for each window size. The figure shows the anomaly detection results (AUC) for each sequence length with five window size values, including sequence-level, window-level (mean), and window-level (max).} \label{fig3}
\end{figure}

This evaluation uses the following performance metrics: accuracy, precision, recall, F1-score, and Area Under the Curve (AUC). Fig.~\ref{fig3} presents the overall performance of \textbf{GUARD-CAN} proposed in this paper.

In addition, the best performance result for each window size is shown in Table~\ref{best_performance_result}. Win and Seq columns represent window size and sequence length, respectively. In addition, the Type column indicates three methods for anomaly detection: sequence, mean and max. 

\begin{table}[htbp]
\centering
\caption{Performance comparison of the best results by window size.}
\label{best_performance_result}
\begin{tabularx}{\textwidth}{
    |c|c|>{\centering\arraybackslash}X|
    >{\centering\arraybackslash}X| >{\centering\arraybackslash}X|
    >{\centering\arraybackslash}X| >{\centering\arraybackslash}X|
    >{\centering\arraybackslash}X|
}
\hline
Win & Seq & Type & Accuracy & Precision & Recall & F1-score & AUC \\
\hline
50  & 50    & sequence       & 0.9702 & 0.9902 & 0.9559 & 0.9727 & 0.9930 \\
    &       & mean           & 0.8682 & 0.9867 & 0.7571 & 0.8544 & 0.9877 \\
    &       & max            & 0.9279 & 0.9557 & 0.9064 & 0.9258 & 0.9803 \\
75  & 50    & sequence       & 0.9588 & 0.9904 & 0.9372 & 0.9630 & 0.9864 \\
    &       & mean           & 0.9342 & 0.9759 & 0.8958 & 0.9340 & 0.9867 \\
    &       & max            & 0.9513 & 0.9389 & 0.9696 & 0.9540 & 0.9697 \\
100 & 120   & sequence       & 0.9578 & 0.9964 & 0.9400 & 0.9673 & 0.9917 \\
    &       & mean           & 0.8870 & 0.9256 & 0.8552 & 0.8854 & 0.9693 \\
    &       & max            & 0.8937 & 0.8874 & 0.9239 & 0.9019 & 0.9509 \\
125 & 50    & sequence       & 0.9645 & 0.9881 & 0.9532 & 0.9700 & 0.9895 \\
    &       & mean           & 0.9035 & 0.9628 & 0.8482 & 0.8987 & 0.9758 \\
    &       & max            & 0.9160 & 0.9483 & 0.8886 & 0.9145 & 0.9713 \\
150 & 120   & sequence       & 0.8762 & 0.9604 & 0.8631 & 0.9079 & 0.9402 \\
    &       & mean           & 0.7542 & 0.9523 & 0.5759 & 0.6066 & 0.9703 \\
    &       & max            & 0.8756 & 0.8713 & 0.9054 & 0.8821 & 0.9309 \\
\hline
\end{tabularx}
\end{table}

\subsection{Result Analysis}
Table~\ref{best_performance_result} shows that the performance of \textbf{GUARD-CAN} tends to decrease as the window size increases. In the experiments, the model achieved 0.9702 accuracy and 0.9930 AUC when the window size was 50 and the sequence length was 50. This result supports our earlier analysis. A larger window size causes the information within each graph to become diluted, which prevents GCN from extracting meaningful features.  Meanwhile, window sizes of 100 and 150 showed their best performance when the sequence length was relatively long. However, the performance remained lower than that of experiments using smaller window and sequence settings. In addition, Fig.~\ref{fig3} further shows that shorter sequence lengths tend to reduce the variation across the three anomaly detection metrics for each window size. These findings indicate that \textbf{GUARD-CAN} can extract more meaningful features and achieve more stable training when using smaller window sizes and sequence lengths.

We also compare model performance when using normalized byte values (float values in the range [0, 1]) and when using binarized byte values (integer values of 0 or 1) as node features. Table~\ref{best_performance_feature_result} presents the top 3 sequence-level anomaly detection results of \textbf{GUARD-CAN} using normalized byte values. The results show that using binarized byte values leads to higher anomaly detection performance. 

Using normalized byte values causes confusion for overcomplete AE and GCN because the value distribution becomes highly dense. In particular, the small differences among normal messages increase the reconstruction error and lead to false positives (FP). In addition, when attack messages such as replay and spoofing resemble the distribution of normal messages, the model can produce false negatives (FN). In addition, normal messages account for the majority of the dataset used in the experiments. This data imbalance remains a major cause of FP and FN even when using binarized byte values. To address this issue, future work will expand the research by applying oversampling techniques focused on anomalous data.

\begin{table}[htbp]
\centering
\caption{Sequence-level anomaly detection performance with normalized byte values. This result shows that the overall performance when normalizing byte values is relatively lower than the case of binarized byte encoding.}
\label{best_performance_feature_result}
\begin{tabularx}{\textwidth}{
    |c|c|>{\centering\arraybackslash}X|
    >{\centering\arraybackslash}X| >{\centering\arraybackslash}X|
    >{\centering\arraybackslash}X| >{\centering\arraybackslash}X|
    >{\centering\arraybackslash}X|
}
\hline
Win & Seq & Type & Accuracy & Precision & Recall & F1-score & AUC \\
\hline
50  & 30    & sequence         & 0.8256 & 0.9254 & 0.7407 & 0.8194 & 0.9225 \\
50  & 50    & sequence         & 0.8095 & 0.9331 & 0.7055 & 0.8031 & 0.9145 \\
50  & 150   & sequence         & 0.8040 & 0.8820 & 0.7992 & 0.8339 & 0.8650 \\
\hline
\end{tabularx}
\end{table}

\section{Conclusion}
In this paper, we proposed \textbf{GUARD-CAN}, a CAN anomaly detection framework that combines graph-based learning and time-series modeling. The proposed model divides CAN messages into windows and converts each window into a graph. After embedding the graphs through GCN, a GRU-based time-series model detects anomalies while reflecting temporal patterns.

The model is designed to effectively learn the structural and temporal characteristics inherent in CAN traffic without complex feature engineering. We also analyzed its performance from multiple perspectives using both sequence-level and window-level evaluation criteria. Furthermore, we demonstrate the importance of selecting an appropriate learning unit through a window size analysis based on Shannon entropy. As a result, \textbf{GUARD-CAN} achieves high detection performance for four representative CAN attacks and especially shows the best performance under small window sizes and short sequence lengths.

In future work, we plan to improve the model by applying a combined loss function that can optimize both sequence-level and window-level detection. In addition, we will also improve \textbf{GUARD-CAN} by modifying the structure to directly use arbitration IDs as node identifiers in GCN learning.

\subsubsection{\ackname}This work was supported by a grant (UI247022TD) from the Agency for Defense Development, South Korea.

\end{document}